# Nitrogen doping induced metal-insulator transition with iso-symmetric character in rutile VO$_2$

Baichen Lin[1,14,15], Shanquan Chen[1,15], Yubo Zhang[2,15], Yangyang Si[1], Haoliang Huang[3], Chuanrui Huo[4], Frans Munnik[5], Yongqi Dong[6], Lu You[7], Jian Shao[1], Yu-Chieh Ku[8], Nguyen Nhat Quyen[8], Aryan Keshri[9], Zhenlin Luo[6], Weiwei Zhao[10], Chun-Fu Chang[11], Chih-Wei Luo[8,12,13], Sujit Das[9], Shiqing Deng[4,*], Chang-Yang Kuo[8,13,**], and Zuhuang Chen[1,16,***]

[1]State Key Laboratory of Precision Welding and Joining of Materials and Structures, School of Materials Science and Engineering, Harbin Institute of Technology, Shenzhen, 518055, China
[2]Minjiang Collaborative Center for Theoretical Physics, College of Physics and Electronic Information Engineering, Minjiang University, Fuzhou, 350300, China
[3]Quantum Science Center of Guangdong-Hong Kong-Macao Greater Bay Area, Shenzhen 518045, China.
[4]Beijing Advanced Innovation Center for Materials Genome Engineering, University of Science and Technology Beijing, Beijing, 100083, P.R. China
[5]Helmholtz-Zentrum Dresden – Rossendorf, Bautzner Landstraße 400, 01328, Dresden, Germany
[6]National Synchrotron Radiation Laboratory, University of Science and Technology of China, Hefei, Anhui 230026, China
[7]School of Physical Science and Technology, Jiangsu Key Laboratory of Thin Films, Soochow University, Suzhou, 215006, China
[8]Department of Electrophysics, National Yang Ming Chiao Tung University, Hsinchu, 30010, Taiwan
[9] Materials Research Centre, Indian Institute of Science, Bangalore 560012, Karnataka, India
[10]Sauvage Laboratory for Smart Materials, The School of Integrated Circuit, Harbin Institute of Technology, Shenzhen 518055, China
[11]Max-Planck Institute for Chemical Physics of Solids, Nöthnitzer Str. 40, 01187 Dresden, Germany
[12]Institute of Physics and Center for Emergent Functional Matter Science, National Yang Ming Chiao Tung University, Hsinchu, Taiwan
[13]National Synchrotron Radiation Research Center, 101 Hsin-Ann Road, Hsinchu 300092, Taiwan
[14]Present address: School of Materials Science and Engineering, Nanyang Technological University, 50 Nanyang Avenue, Singapore 639798, Singapore
[15]These authors contributed equally
[16]Lead contact
*Correspondence: sqdeng@ustb.edu.cn
**Correspondence: changyangkuo@nycu.edu.tw
***Correspondence: zuhuang@hit.edu.cn

**SUMMARY**

Metal-insulator transitions (MITs) in correlated oxides offer immense potential for next-generation Mottronic devices. However, their integration into practical applications is often hindered by the coupling of MITs with symmetry-lowering structural phase transitions, which limits switching speed and endurance. In this study, we engineered an iso-symmetric MIT on average in epitaxial rutile $VO_2$ thin films *via* an *in-situ* nitrogen doping strategy. Nitrogen incorporation effectively suppresses V-V dimerization, enabling an iso-symmetric MIT, while preserving the original crystal symmetry. Furthermore, *in-operando* time-resolved optical reflectivity measurements revealed a shortened switching time in nitrogen-doped films, highlighting their enhanced performance. Our findings provide critical insights into the underlying mechanisms of MITs and introduce anion doping as a powerful tool for tailoring phase transitions in strongly correlated electron systems. This approach opens new avenues for the development of high-performance electronic and photonic devices.

**KEYWORDS**

Metal-insulator transition, iso-symmetric phase transition, nitrogen doping, $VO_2$ thin film

## INTRODUCTION

The intricate interplay of spin, charge, orbital, and lattice degrees of freedom in strongly correlated electronic systems forms the foundation of their extraordinary electronic complexity, driving phenomena such as high-temperature superconductivity[1], giant magnetoresistance[2], and the metal-insulator transitions (MITs)[3, 4]. Among these, materials exhibiting MIT are particularly compelling, offering dynamic, field-controlled switching capabilities across electrical, optical, and magnetic domains. This versatility positions them as cornerstone materials for emerging Mottronics devices, heralding a new era of advanced electronic technologies[5, 6].

Bulk vanadium dioxide ($VO_2$) stands as a prototypical correlated material, renowned for its MIT occurring near 340 K. Traditionally, this MIT is tightly coupled to a structural phase transition (SPT), where the high-temperature tetragonal rutile structure (*R*, space group *P*4$_2$/*mnm*) transforms into a low-temperature monoclinic structure with vanadium-vanadium (V-V) dimers (*M*1, space group *P*2$_1$/*c*) (**Figure 1A**, left panel). Crucially, the temporal bottleneck in this SPT stems not from local atomic displacements but rather from the long-range symmetry rearrangements[7]. Since the critical MIT temperature ($T_{MIT}$) in $VO_2$ is close to room temperature, $VO_2$ becomes a prime candidate for various emerging technologies, including ultrafast photonic and electronic switches[8, 9], steep-field-effect transistors[10], memory storage devices[11, 12], and neuromorphic circuits[13]. Despite its potential, the symmetry-driven monoclinic shearing associated with the SPT imposes significant limitations on switching speed[7] and device endurance[14], presenting a critical hurdle for high-performance applications. Moreover, the entangling of atomic and electronic structures complicates the understanding of the underlying MIT mechanism in $VO_2$, fueling ongoing debate over whether the primary mechanism is governed by electron-lattice interactions (Peierls transition)[15], electron-electron interactions (Mott transition)[16], or a combination of both (Mott-Peierls transition)[17]. This fundamental ambiguity has fueled extensive research, as $VO_2$ not only offers immense potential for ultrafast switching applications but also serves as a model system for exploring correlated electron phenomena. To advance both application and theory, it is of paramount interest to decouple the Mott transition from Peierls distortion, enabling the realization of a MIT coupled with an iso-symmetric structural phase transition (ISPT).

To address this challenge, innovative approaches such as interface engineering[18] and redox gating[19], have been developed to construct and probe iso-symmetric MITs, along with the development of advanced ultrafast photon-based[20-23] and electron-based[7, 9, 24, 25] detection technologies, now enable the precise monitoring of SPT on *ps* and even *fs* scales. For example, the metallic monoclinic phase in $VO_2$ has been reported, induced by heterointerface-induced[18] or photoexcited[21-23] hole carriers (**Figure 1A**, middle panel). However, these metallic monoclinic states are typically transient, manifesting as rutile/monoclinic heterostructures or metastable, photon-excited phases. Despite these advances, achieving a stable, iso-symmetric, single-phase MIT in $VO_2$ that seamlessly correlates with its electrical properties remains an unfulfilled goal.

In this study, we demonstrate a groundbreaking iso-symmetric MIT in nitrogen-doped $VO_2$ (N-doped $VO_2$) epitaxial thin films, maintaining a stable rutile structure with tetragonal symmetry (**Figure 1A**, right panel), indicative of a pure Mott transition mechanism. Remarkably, nitrogen doping effectively suppresses V-V dimer formation, which not only preserves the structural symmetry but also dramatically enhances the MIT speed compared to undoped $VO_2$, paving the way for ultrafast electronic applications.

The SPT in $VO_2$, from the high-temperature rutile to the low-temperature monoclinic phase, is predominantly driven by V-V dimerization. Previous theoretical study[26] and our DFT calculations (Figure S1), imply that hole carriers play notable roles in weakening the V-V dimer structure. This insight guides the design of a stable insulating rutile phase through strategic hole doping. To achieve this, nitrogen was chosen as the substitutional doping source, as its ionic radius closely matches that of $O^{2-}$ (**Figure 1B**), ensuring compatibility with the $VO_2$ lattice. Conventional hole-doping methods, such as low-energy $N_2^+$ ion sputtering[27] and metallic cation doping[28], often struggle to produce high-quality samples. These approaches frequently introduce excessive oxygen vacancies and induce significant structural distortions, which obscure the intrinsic effects of hole doping. To overcome these challenges, we employed the reactive pulsed laser deposition (PLD) technique[29, 30]. This advanced technique enables precise *in-situ* nitrogen incorporation into the VO$_2$ lattice under controlled nitrogen and oxygen atmospheres, facilitating the growth of high-quality epitaxial thin films with minimal structural disruption.

## RESULTS

### Epitaxial growth of high-quality N-doped VO$_2$ thin films

The pure and N-doped VO$_2$ thin films were epitaxially grown on (001)-oriented TiO$_2$ single-crystal substrates using reactive PLD. Nitrogen was introduced during growth by adjusting the atmosphere through controlled N$_2$ and O$_2$ gas flow rates, maintaining an optimized N$_2$/O$_2$ gas flow ratio of 1:1 for the N-doped VO$_2$ samples (Figure S2). Room-temperature X-ray diffraction (XRD) patterns confirmed the single-phase nature of all films, with no detectable impurities (**Figure 1C**). For the pure VO$_2$ film, the diffraction peak at 65.2°, corresponding to the $(\bar{4}02)_m$ reflection associated with the monoclinic structure. In contrast, the diffraction peak for the N-doped film shifts to 65.8°, consistent with the $(002)_t$ reflection of the rutile structure[31]. This indicates a nitrogen doping-induced structural phase transition to the rutile phase at room temperature. Selected-area electron diffraction (SAED) reveals the characteristic half-order reflections of the monoclinic phase, also confirming that pure VO$_2$ film is monoclinic at room temperature, whereas N-doped VO$_2$ film adopts the rutile structure (Figure S3). Moreover, the presence of pronounced Laue oscillations around the $(002)_t$ diffraction peak in the N-doped VO$_2$ thin film suggests its high crystallinity and smooth interfaces. The XRD rocking curve analysis further supports this, revealing that the full-width at half-maximum (FWHM) of the N-doped VO$_2$ film is more than five times narrower than that of pure VO$_2$, approaching the FWHM of the single-crystal TiO$_2$ substrate (inset of **Figure 1C**). To examine the microstructure of the film, atomically resolved high-angle annular dark-field (HAADF) images were taken using an aberration-corrected scanning transmission electron microscope (STEM) at 300 K. As shown in **Figure 1D**, the N-doped VO$_2$ thin film exhibits excellent epitaxial growth with high single crystallinity, featuring an atomically sharp and coherent interface with the TiO$_2$ substrate. No dislocations or visible defects were observed, indicating the structural integrity of the film. The vertical alignment of V atom columns with Ti atom columns confirms the coherently strained state of the film. The corresponding convergent beam electron diffraction (CBED) patterns (inset of **Figure 1D**), further validate the rutile structural characteristics of the N-doped VO$_2$ thin film at 300 K. Complementary atomic force microscopy (AFM) scans corroborate the film's high quality, revealing an ultra-smooth surface with a root-mean-square (RMS) roughness of ~210 pm (Figure S4).

X-ray photoelectron spectroscopy (XPS) was employed to evaluate the chemical valence states and nitrogen doping concentration in the VO$_2$ films. As shown in **Figure 1E**, the N 1$s$ binding energy of the N-doped VO$_2$ film appears at ~396.9 eV, indicative of substitutional N$^{3-}$ ions replacing O$^{2-}$ site in the host lattice and forming V-N bonding[32, 33]. Notably, no observable signals corresponding to adsorbed nitrogen species (~401.0 eV)[34] or interstitial N (~400.5 eV)[33] were detected, confirming the successful incorporation of nitrogen into the oxygen lattice. The introduction of nitrogen doping induces hole carriers, resulting in an increased valence state of vanadium. This effect is evidenced by a subtle shift (~0.2 eV) of the V 2$p$ peaks towards higher binding energies in the N-doped VO$_2$ film compared to the undoped film (right panel of **Figure 1E**). These observations are further corroborated by X-ray absorption spectroscopy (XAS) measurements (Figure S5), which support the role of nitrogen doping in modulating the electronic structure of the VO$_2$ films. In contrast, previous studies on N-doped VO$_2$ nanoparticles synthesized by the hydrothermal method[33] and N-doped VO$_2$ films modified by the surface adsorption method[35] have reported a decrease in the V 2$p$ binding energies, corresponding to a reduced V valence. This reduction is likely due to the formation of oxygen vacancies during sample fabrication, as demonstrated in prior studies[36] and confirmed by our XPS experiments (Figure S2). Such results further confirm the high quality of our thin films, with negligible oxygen vacancy generation upon N doping. Using the relative sensitivity factor method of XPS, the concentration of substitutional N$^{3-}$ in the N-doped VO$_2$ thin film was determined to be ~1% (Figure S2), which was further confirmed by the elemental depth profile obtained through elastic recoil detection analysis (EDRA) using the program NDF v9.3g[37] (**Figure 1F** and Figure S6). Moreover, these depth profile results also indicate a bulk-doping effect, showing the feasibility of our *in-situ* doping strategy. The hole-like character of the charge carriers (i.e., p-type) in the N-doped VO$_2$ films is further confirmed by Hall effect measurements, which yields a positive Hall coefficient of ~6.6 × 10$^{-9}$ m$^3$/C. The corresponding hole carrier density is estimated to be ~9.4 × 10$^{20}$ cm$^{-3}$ (Figure S7), in close agreement with the theoretical value (~1 × 10$^{21}$ cm$^{-3}$) derived from the atomic concentration of N$^{3-}$ ions, corresponding to approximately three hole carriers per 100 VO$_2$ rutile unit cells. Despite the low doping concentration, the resulting hole carrier density is substantial and is expected to play a dominant role in modulating the electrical properties of the films. Additionally, the minimal ionic size difference between nitrogen and oxygen ensures that lattice distortion due to hetero-doping remains insignificant, maintaining the structural integrity of the thin films. This balance between effective doping and preserved crystallinity highlights the potential of our approach for advancing the study of correlated electron systems.

### *On-average iso-symmetric structural phase transition in the N-doped VO$_2$ thin films*

To evaluate the impact of nitrogen doping on the MIT and SPT in VO$_2$ thin films, we conducted detailed electrical and structural characterizations. **Figure 2A** shows resistivity-temperature curves of the N-doped VO$_2$ thin films with varying thicknesses, where all N-doped films exhibit a lower MIT temperature ($T_{MIT}$) compared to that of bulk VO$_2$[38]. This result is consistent with the hole doping effect based on previous theoretical simulations[39]. Interestingly, a consistent $T_{MIT}$ of ~280 K is observed in films with thicknesses below 45 nm, indicating a thickness-independent phase transition behavior in this regime. In contrast, the 70 nm-thick N-doped VO$_2$ films show a higher $T_{MIT}$ of ~310 K and display two distinct transition stages. Given the well-established modulation of the transition temperature by strain and doping strategies [28, 31, 40-42], this behavior likely arises from strain relaxation in regions farther from the substrate, resulting in the coexistence of multiple phases with different strain states in thicker films. This partial strain relaxation is also corroborated by our RSM results (Figure S8). To investigate the SPT characteristics, we selected a 25-nm-thick N-doped VO$_2$ film as a proof-of-concept sample and compared its behavior with that of a 25-nm-thick pure VO$_2$ film. To probe the strain state and lattice symmetry, we performed XRD reciprocal space mappings (RSM) around the (002), (112), and (202) reflections of the TiO$_2$ substrate at room temperature (**Figure 2B**). In all RSM studies, the Bragg peaks of the N-doped VO$_2$ film were aligned with those of the substrate along the in-plane ($Q_x$-value), suggesting that the film is coherently strained to the substrate. Notably, the (002), (112), and (202) reflections also align along the out-of-plane direction ($Q_z$-value), indicating that the N-doped VO$_2$ film essentially exhibits tetragonal symmetry at room temperature. The lattice parameters of the N-doped VO$_2$ film are *a* = *b* = 4.60 Å and *c* = 2.84 Å, which are distinctly different from the monoclinic structure of pure VO$_2$ films (Figure S9). Further, temperature-dependent XRD *θ-2θ* scans were also performed to monitor the possible SPT in the N-doped VO$_2$ thin films (**Figure 2C**). Upon cooling, the out-of-plane lattice parameter shows a noticeable increase around 280 K, aligning with the $T_{MIT}$ observed in electrical transport measurements (**Figure 2A**). Besides, the diffraction pattern exhibits a single sharp reflection, as confirmed by the off-axis reciprocal space mapping (RSM) results at 250 K (Figure S10). This observation suggests that, during the MIT, the in-plane lattice remains barely changed before and after the phase transition if taking no account of thermal expansion, indicating a volume expansion occurs at the MIT. Similar volume changes have also been observed in photoinduced MIT in various Mott insulators with iso-symmetric transitions[43, 44].

The symmetry evolution during the SPT in both pure and N-doped VO$_2$ thin films was further investigated using temperature-dependent Raman spectroscopy. A 325 nm wavelength UV-laser was employed to minimize interference from the substrate signal[18]. At 300 K, the pure VO$_2$ thin film with a monoclinic *P*2$_1$/*c* structure (a subgroup of 2/*m*) displays nine symmetric ($A_g$) and nine antisymmetric ($B_g$) vibration modes[45], serving as its identifiable structural fingerprints (**Figure 2D**). Notably, the Raman peak at 196 cm$^{-1}$ in the pure VO$_2$ thin film represents one of the $A_g$ vibration modes in the 2/*m* point group[45]. In contrast, the N-doped VO$_2$ thin film at 300 K only exhibits a weak and broad Raman signal, which is attributed to phonon damping in the rutile structure[45]. When the pure VO$_2$ film is cooled down from 370 K to 250 K, it undergoes a typical monoclinic distortion, as evidenced by the appearance of structural Raman features associated with the monoclinic phase once the temperature drops below its structural phase transition temperature ($T_{SPT}$) (**Figure 2E**). In contrast, the N-doped VO$_2$ thin film shows no structural fingerprints of monoclinic distortions across the entire tested temperature range, as evidenced by the Raman spectra (**Figure 2F**). This observation suggests the absence of V-V structural dimers, even at temperatures below $T_{SPT}$, thereby indicating that the rutile structure remains stable down to 250 K. To further verify the presence of the low-temperature rutile phase, we performed temperature-dependent four-dimensional STEM (4D-STEM) experiments and extracted CBED patterns (**Figure 2G** and Figure S11). No monoclinic distortions were observed as the sample was cooled down below $T_{SPT}$. Specifically, the diffraction pattern of the N-doped VO$_2$ thin film along the [010]$_t$ zone axis at 250 K (**Figure 2G**) exhibits the same rutile structural features as those observed at room temperature (inset in **Figure 1D**). Notably, the reflections indicative of monoclinic distortions (highlighted by two dashed circles in **Figure 2G**) are absent, indicating the retention of the rutile phase[18]. This further confirms that crystal symmetry remains unchanged during the SPT. Note that the structural characterization experiments indicate that the N-doped VO$_2$ film in the low-temperature phase exhibits an on-average rutile structure. Whether monoclinic distortions exist locally will be further investigated using synchrotron X-ray diffuse scattering techniques[46, 47] in future work.

To clarify the phase transition behavior, we compared the temperature-dependent MIT and SPT of pure VO$_2$ and N-doped VO$_2$ thin films (**Figs. 3A and 3B**). In the pure VO$_2$ films (**Figure 3A**), the MIT and SPT occur concurrently below 320 K, characterized by a transition from the rutile to the monoclinic structure. This simultaneous structural and electronic transformation reflects a conventional symmetry-lowering MIT, where the insulating phase emerges as a result of the structural distortion. In stark contrast, the 25 nm-thick N-doped VO$_2$ thin film undergoes an SPT characterized solely by changes in the *c* lattice constant, while maintaining its original crystal symmetry throughout the phase transition (**Figure 3B**). This phenomenon represents a distinct type of SPT without symmetry-breaking, referred to as an iso-symmetric structural phase transition (ISPT). Besides, the electrical transition magnitude in N-doped VO$_2$ film is lower than that of VO$_2$ film, indicating a smaller gap opening size during the MIT. Based on the quantitative analysis of the ISPT and the temperature-dependent electrical resistivity results in the N-doped VO$_2$ film, we propose the following scenario. The N-doped VO$_2$ film undergoes a coupled MIT and ISPT, characterized by a volume change (manifested as variations only in the *c* parameter) due to variations in V-V bond length while maintaining tetragonal symmetry. In this case, below $T_{MIT}$, an insulating rutile phase emerges, distinctly different from the behavior observed in pure VO$_2$ films.

We performed *in-operando* time-resolved reflectivity measurements to evaluate and compare the phase transition speed in pure VO$_2$ and N-doped VO$_2$ thin films. A laser pulse with a photon energy of 3.1 eV was utilized to excite both films and the corresponding photoinduced transient reflectivity change ($\Delta R/R$) spectra during both cooling and warming processes are presented in Figure S12. Upon excitation with a photon energy of 3.1 eV and a fluence of 192 μJ cm$^{-2}$, which is well below the excitation threshold[48-51], the $\Delta R/R$ signal rapidly reaches its peak and decays gradually during the recovery process. This decay is overlaid by phonon-induced oscillations. The recovery process, which spans several hundred picoseconds, eventually returns the material to its initial state. **Figure 3C** shows the peak amplitude of $\Delta R/R$ spectra at 0 *ps* (see Figure S12) for pure and N-doped VO$_2$ under both cooling and warming cycles. The intersection of the two dashed lines indicates the MIT onset temperature. It is observed that the metallic phase exhibits a positive $\Delta R/R$, while the insulating phase shows a negative $\Delta R/R$. Near the onset temperature, laser pulses can induce a phase transition from insulating to metallic states in VO$_2$. This is evident from the change in the sign of $\Delta R/R$ over time after the system is pumped by the laser. **Figure 3D** (pure VO$_2$) and **Figure 3E** (N-doped VO$_2$) present the time-dependent $\Delta R/R$ spectra at the onset (blue/red curves) and their neighboring temperatures (gray curves). Notably, the return time to the initial insulating state (negative $\Delta R/R$), indicated by red arrows, is shorter for N-doped VO$_2$ (~207 *ps* at 264 K) compared to pure VO$_2$ (~264 *ps* at 310 K). This demonstrates that the ISPT in N-doped VO$_2$ exhibits a faster MIT switching speed.

### *Orbital assisted MIT in the N-doped VO$_2$ thin films*
The iso-symmetric phase transition suggests that the MIT in N-doped VO$_2$ is primarily governed by electronic factors. In VO$_2$, where V is coordinated by octahedral oxygen, the *d* orbitals are split into three $t_{2g}$ orbitals and two $e_g$ orbitals. The $t_{2g}$ orbitals comprise $d_{x^2-y^2}, d_{xz}$, and $d_{yz}$ orbitals. Among them, the $d_{x^2-y^2}$ orbital overlaps with the neighboring V atoms in the V-V chains, forming the $d_\parallel$ band, while the $d_{xz}$ and $d_{yz}$ orbitals contribute to the $\pi^*$ band (**Figure 4A**, upper panel). In conventional VO$_2$, orbital polarization is recognized to play a key role in assisting the MIT[17]. Thus, it is crucial to assess its influence on the distinct MIT behaviors observed in our N-doped films. Using density-functional theory (DFT) calculations, we simulated both the zero- and high-temperature phases of the rutile structure, with and without spin-polarization. As shown in **Figure 4B**, the high-temperature phase remains metallic as the three $t_{2g}$ orbitals are partially populated, each holding roughly one-third of an electron. In contrast, the zero-temperature phase develops a finite bandgap as the $d_{x^2-y^2}$ orbital repopulates itself with an entire electron, breaking the $t_{2g}$ orbital isotropy without any structural dimerization (**Figure 4C**). This polarization of the $d_{x^2-y^2}$ orbital is directly visualized through electron redistribution, where electrons accumulate in this orbital during the MIT (**Figure 4A**, bottom panel). To address inter-electron correlation more accurately, we also conducted single-site dynamical mean-field theory (DMFT) simulations[52]. The results further support our findings: rutile VO$_2$ remains metallic at 350 K (**Figure 4D**) but exhibits a bandgap with pronounced orbital polarization at 250 K (**Figure 4E**). The emergence of a bandgap in the current single-site DMFT suggests that the gapping mechanism in rutile VO$_2$ is akin to the conventional Mott mechanism[53], in contrast to the dimer-based mechanism, which would require cluster DMFT[54]. Experimentally, anisotropic orbital occupation can lead to discrepant responses in polarized X-ray, resulting in noticeable variations in the total electron yield (TEY) signal in polarization-dependent XAS. In this sense, the X-ray linear dichroism (XLD) spectrum, obtained from the difference between two XAS spectra measured with orthogonal polarizations ($I_\perp$ and $I_\parallel$), effectively captures these orbital effects (Figure S13). To verify our theoretical findings, we performed polarization-dependent XAS measurements along V $L$-edge and O $K$-edge on pure and N-doped VO$_2$ films, at temperatures of 350 K (metallic state) and 250 K (insulating state), corresponding to the phase transitions in both samples. For pure VO$_2$ films, the orbital occupation difference, derived from dichroic signal changes in metallic and insulating states, indicates the electronic structure transition (Figure S14). This behavior is consistent with previous studies on both bulk and thin-film VO$_2$[17, 41, 55], where the transition is accompanied by symmetry breaking as the material shifts from the metallic to insulating state. In contrast, for N-doped VO$_2$ films at 350 K (**Figure 4F**), the TEY signals collected from two polarized directions are similar, resulting in minimal dichroism, indicating isotropic orbital occupation in the metallic state. At 250 K, however, an enhanced dichroism at ~514.7 eV was observed, suggesting selective filling of the $d_\parallel$ orbital (**Figure 4G**), which is consistent with our DFT theoretical simulations.

The anisotropic occupation observed in the insulating VO$_2$ is conventionally attributed to the orbital polarization caused by V-V dimerization[41]. However, our structural characterizations of N-doped VO$_2$ films revealed no evidence of V-V dimerization, effectively excluding the possibility of a Peierls transition as the driving mechanism. This strongly suggests that the MIT in N-doped VO$_2$ thin film is driven purely by electron-electron correlations. These findings confirm that N-doped VO$_2$ films undergo an ISPT, although their electronic structure transition still resembles the behaviors observed in the undoped VO$_2$ counterparts. Additionally, the $\pi^*$ orbital and $d_\parallel$ orbital of insulating VO$_2$ can be further probed using XAS O $K$-edge measurements[55], where the $d_\parallel$ orbital can only be detected in $I_\parallel$ due to the $z$ symmetry nature of the O $2p$ orbital. The band gap between the $\pi^*$ and $d_\parallel$ orbitals derived from the O $K$-edge of N-doped VO$_2$ thin film is ~0.89 eV (Figure S15), which is smaller than that of pure VO$_2$ (~0.96 eV) (Figure S14). This reduction can result from the N dopant-induced hole doping effect, which generates an intermediate energy level, consequently reducing the band gap and electric resistivity. Previous first-principles calculations suggest that the bandgap reduction induced by hole doping may result from orbital switching between the $d_{x^2-y^2}$ and $d_{xz}/d_{yz}$ orbitals.[39] Our XLD measurements are consistent with such orbital switching. The structural pinning and reduction of bandgap reduction contribute to metallicity, in agreement with prior DFT calculations.[56, 57] This is consistent with the reduced electrical transition magnitude observed in the N-doped VO$_2$ films compared to the pure VO$_2$ films (**Figures 3A and 3B**).

## DISCUSSION

There has been a long-standing debate regarding the origin of the MIT in $VO_2$ thin films: whether it is driven primarily by a Peierls transition or by a Mott-Hubbard transition (electron-electron correlation). In this work, we provide solid experimental and theoretical evidence for an isosymmetric MIT in N-doped $VO_2$ thin films, which occurs without Peierls distortion. More importantly, a non-transient insulating rutile structural phase is achieved via anion doping, in contrast to the photoinduced or interface-induced metallic monoclinic phase reported in previous studies[18, 22]. Our findings underscore the pivotal role of N-doping-induced hole carriers in enabling the iso-symmetric MIT. Photoinduced iso-symmetric insulator-metal transition experiments in metastable monoclinic $VO_2$ corroborate the significance of hole carriers in altering macroscopic electronic properties without breaking structural symmetry[21-23]. Previous studies have shown that the introduction of hole carriers generally tends to collapse the band gap and enhance metallicity[21, 22, 58]. Here, our results suggest that hole carriers with proper concentration primarily suppress V-V dimerization by slightly reducing the Coulomb repulsion[21], thereby enabling an iso-symmetric phase transition. At low temperature, however, the Coulomb repulsion remains sufficiently strong to localize the electrons, opening the bandgap and preserving the insulating state. Similar insulating rutile phases have previously been observed only observed in ultrathin undoped $VO_2$ films with thicknesses below 10 nm[59, 60]. Consistent with this observation, our DFT and DMFT calculations show a band-gap opening in insulating rutile $VO_2$ at low temperature (Figures 4C and 4E). Furthermore, the minimal hole carrier density required for this transition is further validated by photo-excitation experiments[21] and theoretical calculations[61], highlighting a fundamental mechanism of electronic control in $VO_2$.

In conclusion, our study unveils an iso-symmetric MIT in high-quality N-doped $VO_2$ thin films, where the introduction of hole carriers lowers $T_{MIT}$ to ~280 K and stabilizes the rutile structure in the insulating phase, suggesting a Mott transition mechanism. Notably, the iso-symmetric SPT enhances the MIT switching speed, offering a critical advantage for ultrafast applications. Although the enhancement is modest compared with pure $VO_2$, it represents the first demonstration that an iso-symmetric phase transition can accelerate MIT kinetics. This finding points to a practical and scalable pathway for optimizing $VO_2$-based switching devices. Furthermore, the transition temperature can be further tuned toward room temperature via strain engineering[41] or by controlling the doping concentration[39], enabling practical device implementation. These findings provide profound insights into the MIT mechanism, paving the way for the development of next-generation Mottronics devices with superior response capabilities and energy-efficient performance.

### METHODS

**Sample preparation.** The epitaxial pure $VO_2$ and N-doped $VO_2$ films were grown on single-crystalline $TiO_2$ (001) substrates by pulsed-laser deposition (Arrayed Materials RP-B) with a KrF excimer laser ($\lambda$ = 248 nm). Both pure $VO_2$ and N-doped $VO_2$ films were grown at 400 °C from a $VO_2$ ceramic target at a laser repetition rate of 10 Hz, and the laser fluence was ~1.1 J cm$^{-2}$. For all films, the growth took place in an on-axis geometry with a target-substrate distance of 5.0 cm. The total atmosphere pressure remained at $20 \pm 0.5$ mTorr, while the gas ratio $N_2:O_2$ was controlled by adjusting the entrance fluence of oxygen and nitrogen. For the N-doped $VO_2$ case, the fluence of $N_2$ and $O_2$ gases is about 4 sccm. The growth rate of epitaxial thin film is about 2.7 nm/min. To ensure that the heterogeneous gases are evenly distributed, it is left for 10 min after the gases are passed into the chamber. After deposition, samples were annealed at 400 °C for 10 min and then cooled to room temperature at a rate of 10 °C/min under the growth atmosphere condition.

**Characterizations**. XRD, RSM, XPS, XAS, XLD, ERDA, Raman spectroscopy, STEM, Time-resolved ultrafast spectroscopy and electrical tests were performed to analyze the crystal structure, electronic structure, chemical composition, and electrical and optical properties. More details can be found in the Supporting Information.

**DFT calculations.** The DFT electronic structure is performed using Quantum ESPRESSO[62, 63]. The exchange-correlation calculations in the main text are based on the recently developed r2SCAN meta-GGA[64]. These are followed by the application of the Wannier90 code [65] to downfold the electronic structure into the $t_{2g}$ orbital subspace, which represents the correlated orbitals. The DFT+DMFT calculations[52] within the single-site approximation are carried out using the Solid_DMFT code[66], which is interfaced with TRIQS/DFT tools[67, 68]. The electronic interactions are modeled using the Kanamori Hamiltonian[69], and the quantum impurity problem is solved using the continuous-time hybridization-expansion Monte Carlo solver[70].

## RESOURCE AVAILABILITY

### *Lead contact*
- Requests for further information and resources should be directed to and will be fulfilled by the lead contact, Zuhuang Chen (zuhuang@hit.edu.cn).

### *Materials availability*
- This study did not generate new unique reagents.

### *Data and code availability*
- Any additional information required to reanalyze the data reported in this paper is available from the lead contact upon request.

**ACKNOWLEDGMENTS**

This work was supported by National Natural Science Foundation of China (Grant Nos. 52525209, 92477129, 52372105 and 22375015) and the Guangdong Basic and Applied Basic Research Foundation (Grant No. 2024B1515120010). Z.H.C. acknowledges the financial support for Outstanding Scientific and Technological Innovation Talents Training Fund in Shenzhen. Parts of this research were carried out at IBC at the Helmholtz-Zentrum Dresden-Rossendorf E.V., a member of the Helmholtz Association. The authors thank the staff from Shanghai Synchrotron Radiation Facility (SSRF) at BL02U2.

**AUTHOR CONTRIBUTIONS**

Z.H.C. and B.C.L. conceived and designed the experiments. Z.H.C. supervised this study. B.C.L. and S.Q.C. fabricated the films and performed the XRD and electrical transport measurements with the assistance of Y.Y.S., H.L.H., J.S. and W.W.Z.. B.C.L. and S.Q.C. performed the temperature-dependent Raman spectrum characterizations with the assistance of L.Y.. S.Q.D. and C.R.H. performed the STEM characterization. Y.B.Z. carried out theoretical calculations. B.C.L. carried out the XPS and AFM measurements. F.M. carried out the ERDA measurements. N.N.Q. and C.W.L. performed the in-operando time-resolved reflectivity measurements. Y.C.K. and C.Y.K. performed the synchrotron XAS characterization with the assistance of C.F.C.. S.Q.C., Y.Y.S., Y.Q.D. and Z.L.L. performed the temperature-dependent synchrotron XRD-RSM characterization. Z.H.C., C.Y.K., Y.B.Z., S.Q.C., S.Q.Deng, S.Das, A.K. and B.C.L. analyzed the data and co-wrote the manuscript. All authors contributed to the discussions and manuscript preparation.

**DECLARATION OF INTERESTS**

Z.H.C. and B.C.L. are inventors on Chinese patent registration [patent No. ZL 2022 1 1083022.X], which covers the in-situ nitrogen-doping method for single crystal oxide thin films described in this manuscript.

**FIGURE (AND SCHEME) TITLES AND LEGENDS**

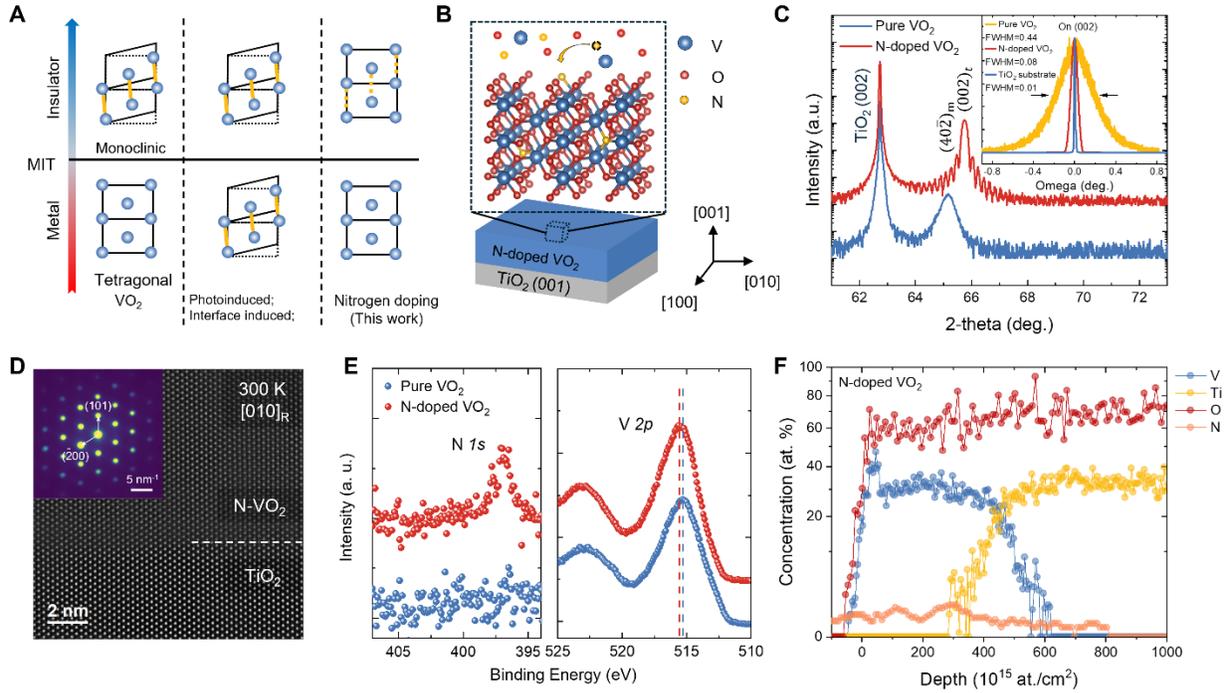

**Figure 1. Fabrication of high-quality N-doped VO$_2$ epitaxial thin films by reactive pulsed-laser deposited technique in N$_2$-containing growth atmosphere.** (A) Strategies for constructing an iso-symmetric phase transition in VO$_2$ films. Left panel illustrates the conventional phase transition from metallic rutile to insulating monoclinic VO$_2$. Middle panel shows the metallic monoclinic phase of VO$_2$ induced by photon excitation or interfacial effects. Right panel demonstrates the rutile insulating phase induced by nitrogen doping. The yellow lines represent the existent (solid) and absent (dashed) V-V dimers, respectively. (B) Schematic illustration of nitrogen doping in rutile VO$_2$, where nitrogen substitutes for oxygen. (C) Room-temperature 2θ-ω scans of pure VO$_2$ and N-doped VO$_2$ thin films. Thickness fringes and a 2θ shift of $(002)_R$ diffraction peaks are observed in the N-doped VO$_2$ thin film. Inset shows XRD rocking curve measurement of the TiO$_2$ substrate, pure VO$_2$, and N-doped VO$_2$ films around the (002) reflection (tetragonal notation). (D) Atomically resolved HAADF-STEM image of N-doped VO$_2$ thin film. Inset shows the corresponding CBED pattern taken at room temperature. (E) XPS of N 1s and V 2p core levels in pure and N-doped VO$_2$ thin films. The V 2p peak shows a subtle shift towards higher binding energy, as highlighted by the dashed lines. (F) Elemental depth profile of the N-doped VO$_2$ thin film.

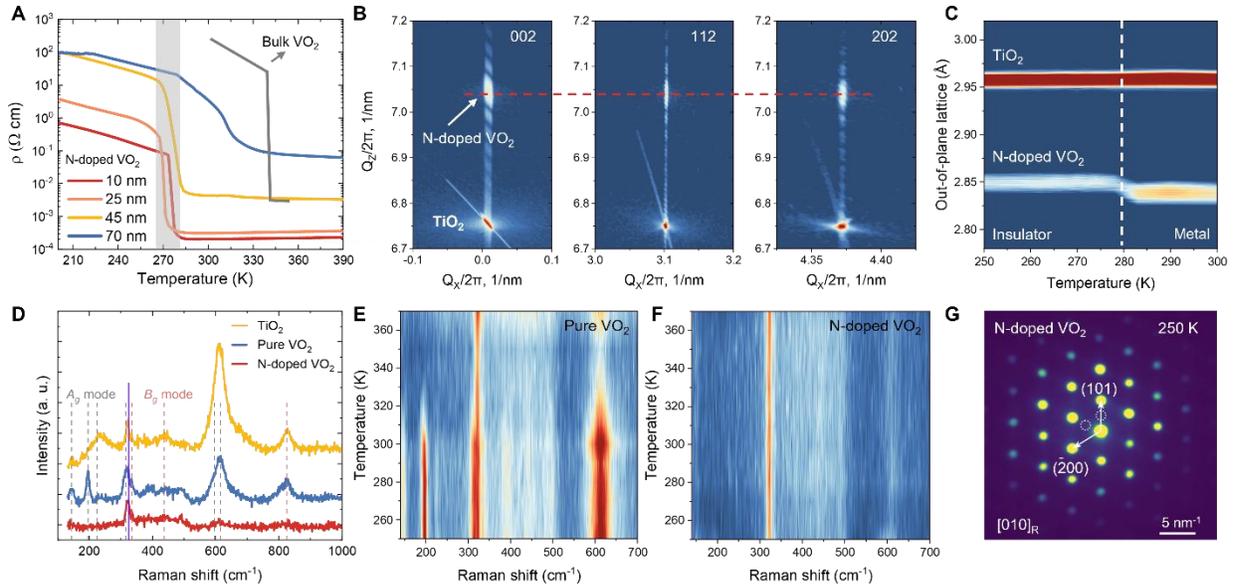

**Figure 2. Iso-symmetric electronic phase transition in 25 nm-thick N-doped VO$_2$ thin films.** (A) Temperature-dependent electrical resistivity of N-doped VO$_2$ thin films with thicknesses of 10 nm, 25 nm, 45 nm, and 70 nm. The grey band highlights the consistent MIT temperature (T$_{MIT}$) for films up to 45 nm. The T$_{MIT}$ of bulk single-crystal VO$_2$ (~340 K) is reproduced from ref. 38 with permission. Copyright 2007, Elsevier. (B) Room-temperature RSM of the N-doped VO$_2$ thin film around the (002), (112), and (202) TiO$_2$ reflections. The (002), (112), and (202) reflection peaks of the N-doped VO$_2$ thin film show the same Q$_z$ value, which is highlighted via a red dashed line. (C) Temperature-dependent XRD 2θ-ω scans of the N-doped VO$_2$ thin film, showing an out-of-plane lattice change at ~280 K. The transition temperature measured by electrical transport is highlighted via a white dashed line for comparison. (D) Room-temperature Raman spectra of the TiO$_2$ substrate, pure VO$_2$, and N-doped VO$_2$ thin films. The A$_g$ and B$_g$ vibration modes of the monoclinic VO$_2$ phase are marked with grey and red vertical dashed lines, respectively. The purple line at ~324 cm$^{-1}$ corresponds to the instrument error resulting from the optical quartz window. (E, F) Temperature-dependent Raman spectrum mapping of (E) pure and (F) N-doped VO$_2$ thin films. No variation in Raman features was observed during the phase transition in the N-doped film. (G) CBED pattern of the N-doped VO$_2$ thin film taken at 250 K.

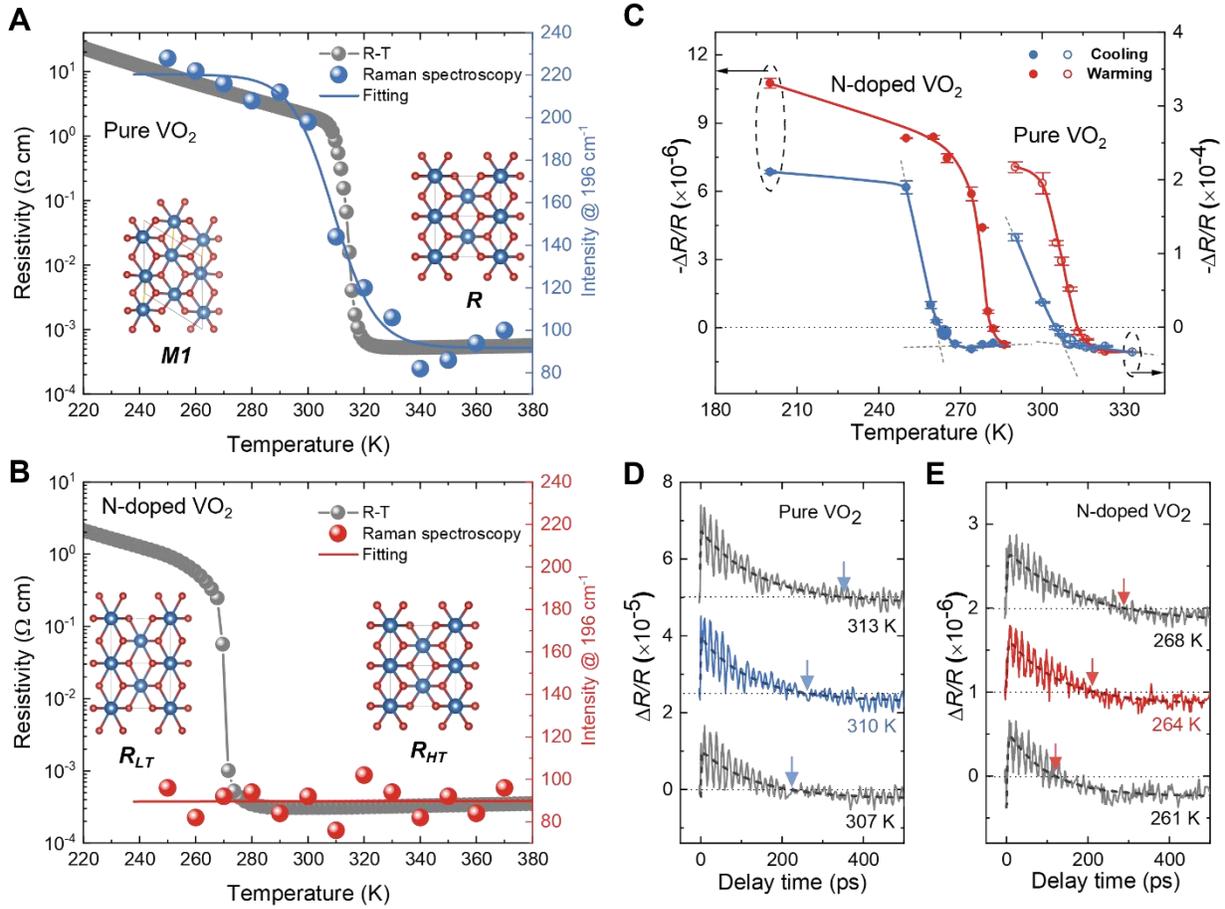

**Figure 3. Enhanced MIT speed in 25 nm-thick N-doped VO$_2$ thin films due to iso-symmetric phase transition.** (A and B) Comparison of the temperature-dependent Raman spectroscopy and resistivity for (A) 25 nm-thick pure VO$_2$ and (B) the 25 nm-thick N-doped VO$_2$ thin films, highlighting the coupling behavior between SPT and MIT. (C to E) Ultrafast dynamics of pure VO$_2$ and N-doped VO$_2$ thin films. (C) Peak amplitude of the transient reflectivity change (ΔR/R) spectra at 0 ps (see Figure S12) of pure VO$_2$ and N-doped VO$_2$ thin films with cooling and warming processes. The -ΔR/R of the N-doped VO$_2$ and pure VO$_2$ thin films are plotted into the left and right Y-axis, respectively, which is indicated by the dashed ovals. Blue and red dots indicate cooling and warming processes, respectively. The linear fitting sections before and after the phase transition are represented by dashed lines. The crossing point of two dashed lines marks the onset temperature (bigger symbol) for MIT. Solid lines serve as visual guide. Error bars represent ± standard deviation. (D and E) ΔR/R spectra of (D) pure and (E) N-doped VO$_2$ thin films at the crossing-point temperature (blue and red, respectively) and their adjacent temperatures in (C). Dashed lines show bi-exponential decay fitting curves. The arrows indicate the time required to return to the initial insulating state at each examined temperature.

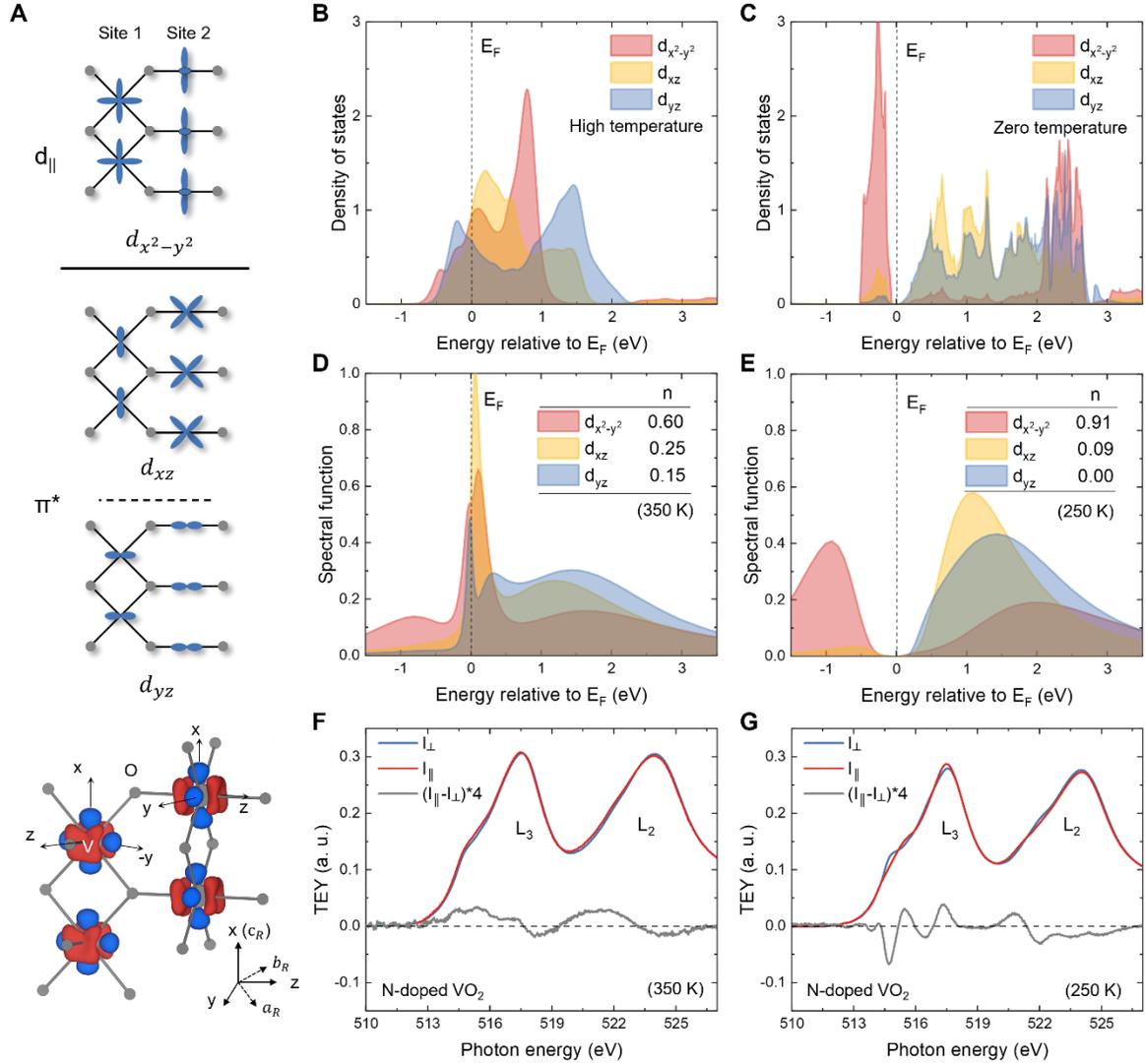

**Figure 4. Orbital-dominated MIT in rutile VO₂.** (A) Schematics of orbital occupation behavior in rutile VO₂. Top: Diagram of t$_{2g}$ orbitals, with $d_{x^2-y^2}$ orbitals forming $d_{||}$ band and $d_{xz}/d_{yz}$ orbitals contributing to $\pi^*$ band. Site 1 is rotated 90° around the c-axis to align with Site 2. Bottom: Local coordinate system defining the d-orbitals on V atoms. Isosurfaces denote the electron redistribution during the metallic-to-insulating transition, expressed as $\Delta n = n^{insulator} - n^{metal}$, where red and blue indicate electron depletion and accumulation, respectively. (B, C) Electronic density-of-states for the (B) high- and (C) zero-temperature phases. (D, E) Spectral functions for phases at (D) 350 K and (E) 250 K, simulated via DMFT. The dashed lines in (B-E) represent the Fermi level. (F, G) XAS spectra of V L$_{2,3}$-edge and XLD for the N-doped VO₂ film measured at (F) 350 K and (G) 250 K. Signal collected using linearly polarized X-rays, where I$_\perp$ and I$_{||}$ represent the electric-field orientation perpendicular and parallel to the rutile c axis, respectively. The zero TEY intensity is highlighted via dashed lines.